\documentclass[journal=jpccck,manuscript=article]{achemso}
\pdfoutput=1

\usepackage{achemso}
\usepackage{amsmath}
\usepackage{url}
\usepackage{tabularx}
\usepackage{graphicx}
\usepackage{subfig}
\usepackage{multirow}
\usepackage{makecell}
\usepackage{upquote}
\usepackage[normalem]{ulem}
\usepackage{dutchcal}

\newcolumntype{Y}{>{\raggedleft\arraybackslash}X}
\newcolumntype{Z}{>{\centering\arraybackslash}X}

\title[]{Structure, Thermodynamics, and Raman Spectroscopy of Rhenium-Doped Bulk MoS$_2$ from First Principles}

\author{Enrique Guerrero}
\email{eguerrero23@ucmerced.edu}
\affiliation{Department of Physics, University of California, Merced, Merced, CA 95343, United States}

\author{David A. Strubbe} \email{dstrubbe@ucmerced.edu}
\affiliation{Department of Physics, University of California, Merced, Merced, CA 95343, United States}

\begin{document}

\begin{abstract}
  Doping MoS$_2$ with Re is known to alter the electronic, structural, and tribological properties. Re-doped MoS$_2$ has been previously mainly studied in monolayer or few-layer form, but can also be relevant for applications in many-layer or bulk form. In this work, we use density functional theory to explore the structure, phase stability, and Raman spectrum of bulk Re-doped MoS$_2$. We consider the possibility of the Re dopant existing at different locations and provide experimentally distinguishable characteristics of the most likely sites: Mo-substitution and tetrahedral (t-) intercalation. We demonstrate and benchmark a general approach to calculate Raman spectra of doped materials with metallic densities of states by using atomic Raman tensors from the pristine material. Applying this method to the metallic Re-doped structures, we find characteristic shifts in the Raman-active peaks depending on Re dopant position: redshifts in both A$_{\rm 1g}$ and E$_{\rm 2g}^1$ peaks in the t-intercalated case versus a redshift for A$_{\rm 1g}$ and blueshift (sometimes accompanied by a smaller redshifted peak) for E$_{\rm 2g}^1$ peaks in the Mo-substituted case, which can be used to identify the dopant sites in experimental samples. We analyze the interactions giving rise to these shifts.
\end{abstract}

\maketitle

	\section{Introduction}
		MoS$_2$ is a versatile semiconductor having an anisotropic, two-dimensional structure, with interesting electronic \cite{Bernardi}, optical, tribological \cite{Vazirisereshk, Rapoport}, catalytic \cite{Mao}, and spintronic properties \cite{Sanikop}. Doping has been used as a strategy to tune these properties for various applications. Re is one of the most-studied MoS$_2$ dopants \cite{Dolui, AlDulaimi, Yoshimura}; since Re has one more $d$-electron than Mo, Re substitution for Mo leads to $n$-type MoS$_2$\cite{Hallam}. It can be incorporated by various synthesis methods \cite{AlDulaimi, Ghoshal, GaoSubst, LinDopant, Yadgarov} and also occurs as a natural impurity in MoS$_2$.\cite{Brandao}
		Most of the recent interest in Re-doped MoS$_2$ systems has been in single-layer structures,\cite{ZhaoDFT, Tian} with few multilayer or bulk studies.\cite{Hallam, AlDulaimi} Studying Re-doping in bulk is nevertheless also interesting, since bulk MoS$_2$ shares some of the opto-electronic properties of single layers, and as an infinite-layer limit helps understand trends with increasing number of layers. Bulk MoS$_2$ is also important in macroscale applications such as solid lubrication \cite{Vazirisereshk,Acikgoz}. 

		MoS$_2$ is an effective solid lubricant owing to the ease of shearing along the basal planes \cite{Vazirisereshk}. Doping MoS$_2$ alters material growth patterns \cite{Kondekar} and enhances tribological properties \cite{Stupp}. Frictional forces on MoS$_2$ have been measured by atomic force microscopy (AFM), and generally friction decreases with more layers, as for other 2D materials \cite{LeeFriction}. In a recent work,\cite{Acikgoz} surprising opposite trends in friction have been observed for Re-doped MoS$_2$ with AFM. In that work, we interpreted the cause of the trend with theoretical calculations on friction and elastic stiffening by the dopants. This work gives background on the basic structural properties of the Re-doped MoS$_2$ used as the basis for those computations.

		The consensus in the literature is that Re in monolayer MoS$_2$ substitutes for Mo, which previous works have indicated with annular dark-field imaging, \cite{GaoSubst, LinDopant} scanning atomic tomography\cite{Tian} experiments, calculations of formation energy,\cite{Dolui} and consideration of the general chemical similarity between Mo and Re. In bulk, neighboring layers allow for the possibility of intercalation. The favored site in bulk is unclear and has not been established by experiments, which are not necessarily able to distinguish between the sites in a multi-layer structures. In particular, there is some experimental evidence of Re intercalation in 2H-MoS$_2$ from Raman spectroscopy interpreted by our density-functional theory (DFT) calculations,\cite{Acikgoz} and in 3R-MoS$_2$ from energy dispersive x-ray analysis on samples from chemical vapor transport \cite{TiongPiez}. Moreover, DFT studies indicate the stability of other transition metals in tetrahedral (t-) intercalation,\cite{Guerrero, Ivanovskaya} meaning dopant sites beyond Mo-substitution are possibilities worth investigating.

		Raman spectroscopy is a key characterization method for 2D materials and can be used to probe differences in microstructure---particularly local bonding configurations.\cite{Guerrero} Doping MoS$_2$ can shift the pristine Raman-active E$_{\rm 2g}^1$ and A$_{\rm 1g}$ peaks. The experimental literature has found different magnitudes and directions of these shifts in monolayer Re-doped MoS$_2$: redshifts in the A$_{\rm 1g}$ peaks,\cite{AlDulaimi, Tian, Ghoshal} redshifts of the E$_{\rm 2g}$ peak,\cite{AlDulaimi, Ghoshal, LiShiSheng, Xia} or blue shifts in the E$_{\rm 2g}$ peak.\cite{Ghoshal} The interpretation of Re-related Raman shifts has been unclear and not investigated in detail, and shifts have generally been vaguely attributed to overall strengthening or weakening of bonds by doping \cite{Iqbal}.
                Calculation of the Raman spectra in such systems poses a problem, because new states exist at the Fermi level compared to the pristine system. The resulting metallic density of states (DOS) cannot be handled by typical DFT methods in the static approximation \cite{Lazzeri}, despite the Raman spectra being physically observable. To overcome this, we develop an method to approximate the Raman tensor and thus obtain a Raman spectrum.

		In this work, our investigation of the structure and bonding of bulk Re-doped MoS$_2$ shows that the phase stability leads us to two structures of interest: the t-intercalated and Mo-substituted Re-doped structures.
        		Our Raman calculations find clear differences between intercalated and Mo-substituted Re-doped MoS$_2$, with redshifts of both active peaks in the intercalated case but a blue-shift of the E$_{\rm 2g}^1$ and red-shift of the A$_{\rm 1g}$ peaks in the Mo-substituted case. These shifts can be experimentally identifiable features. To accomplish these Raman computations, we propose a method to estimate the Raman tensor for a metallic doped system, and benchmark its accuracy.

	\section{Methods}
	We use plane-wave DFT and density functional perturbation theory\cite{Baroni} (DFPT) implemented in \texttt{Quantum ESPRESSO}\cite{QE, QE2009} version 6.6. Calculations were performed using either the Perdew-Burke-Ernzerhof\cite{PBE} (PBE) generalized gradient approximation with the Grimme-D2 (GD2)\cite{GD2} van der Waals correction, or the Perdew-Wang\cite{PW} local density approximation (LDA) (without a van der Waals correction). We use Optimized Norm-Conserving Vanderbilt pseudopotentials\cite{ONCV} parametrized by PseudoDojo\cite{PseudoDojo}.

                PBE+GD2 yields slightly better structural parameters than LDA.\cite{Guerrero} Unlike PBE, however, LDA is compatible with \texttt{Quantum ESPRESSO}'s implementation of Raman intensities, so we decided to use LDA for vibrational spectra (including initial structure optimization) and use PBE+GD2 otherwise. 60 Ry was used as the kinetic energy cutoff for PBE while 80 Ry was used for LDA---a higher cutoff was required for more accurate phonon modes in DFPT but not required for reasonable structural optimization. We applied 0.001 Ry Gaussian smearing to the electronic occupations to be able to handle metallic structures. Electronic energy, force, and pressure thresholds of $10^{-6}$ Ry, $10^{-4}$ Ry/Bohr, and 0.005 GPa were used respectively in variable-cell relaxations.

		We model Re-doped MoS$_2$ with charge-neutral periodic supercells of 2H-MoS$_2$. We use varying supercell sizes and corresponding Monkhorst-Pack $k$-grids to test the dependence on Re concentration and approach the low-doping limit, as summarized in Table \ref{tab:supercell}.
                We tested the effect of out-of-plane dopant interactions on some properties using the $2\times2\times2$ supercell. Electronic DOS calculations used 0.1 eV broadening, and 20\% additional unoccupied states. All $k$-grids are half-shifted. An exception is the $3\times3\times1$ t-intercalated structure which was numerically troublesome and used 15\% unoccupied states and a $\Gamma$-centered grid to converge successfully.

		Re doping introduces an odd number of electrons leading to the possibility of magnetization, and thus we use spin-polarized calculations for relaxations. Magnetization has been observed\cite{Xia} in 1T Re-doped MoS$_2$ and predicted in monolayers.\cite{ZhaoDFT}
                However, we found that energy differences between spin-polarized and spin-unpolarized states are only roughly $10^{-5}$ eV per atom (Table \ref{tab:GD2Mag}), so we conclude that magnetism is not significant in this system. Moreover, our computational treatment of Re-doped MoS$_2$ as a periodic supercell likely overestimates the magnetic effects compared to the probable disordered Re distributions in a real sample. Because of this, we included magnetization when it was feasible (relaxations and total energy for phase diagrams) but used spin-unpolarized calculations for vibrational spectra, which are sensitive to small energy differences.

\begin{table*}[t]
	\setlength{\tabcolsep}{6pt}
	\renewcommand{\arraystretch}{1.5}
	\caption{Supercell and Monkhorst-Pack $k$-grid parameters.}
	\label{tab:supercell}

	\begin{tabular}{| c | c | c | c | c | c| }
		\hline
		supercell & Re concentration & scf $k$-grid & DOS $k$-grid \\
		\hline \hline
		$2\times 2\times 1$ & 4.17 at-\% & $6\times6\times2$ & $8\times8\times4$ \\
		\hline
		$2\times 2\times 2$ & 2.08 at-\% & $6\times6\times1$ & $8\times8\times2$ \\
		\hline
		$3\times 3\times 1$ & 1.85 at-\% & $3\times3\times2$ & $6\times6\times4$ \\
		\hline
		$4\times 4\times 1$ & 1.04 at-\% & $2\times2\times2$ & - \\
		\hline
	\end{tabular}
\end{table*}

		\begin{table*}[t]
			\setlength{\tabcolsep}{6pt}
			\renewcommand{\arraystretch}{1.5}
			\begin{tabular}{| c || c | c || c | c || c | c || c | c || c | c|}
				\hline
				supercell &
				\multicolumn{2}{c|}{Mo subst.} &
				\multicolumn{2}{c|}{S subst. / S vac. + Re} &
				\multicolumn{2}{c|}{o-intercal.} &
				\multicolumn{2}{c|}{t-intercal.} &
				\multicolumn{2}{c|}{Re-Mo interst.} \\
				\hline
				 & $\mu$ & $\Delta E$
				 & $\mu$ & $\Delta E$
				 & $\mu$ & $\Delta E$
				 & $\mu$ & $\Delta E$
				 & $\mu$ & $\Delta E$ \\
				\hline \hline
				$2\times2\times1$ & 0.00 & 0
				& 0.67 & 5.5 & 0.00 & 0
				& 0.00 & 0 & 0.33 & 4.4 \\
				\hline
				$2\times2\times2$ & 0.00 & 0
				& 0.00 & 0 & 0.00 & 0
				& 0.00 & 0 & 0.00 & 0 \\
				\hline
				$3\times3\times1$ & 0.33 & 3.1
				& 3.00 & 84.2 & 1.00 & 4.1
				& 0.00 & 0 & 1.00 & 45.9 \\
				\hline
				$4\times4\times1$ & 0.00 & 0
				& 3.00 & 116.0 & 0.00 & 0
				& 1.00 & 18.0 & 1.00 & 18.0 \\
				\hline
			\end{tabular}
			\caption{Magnetization for Re-doped MoS$_2$ supercells from PBE+GD2, and energy differences between the spin-polarized and spin-unpolarized states.\textsuperscript{\emph{a}}}
                        \textsuperscript{\emph{a}}Magnetization $\mu$ in $\mu_{\rm B}$ and energy difference $\Delta E$ in meV, both per Re atom (one per cell); structures as defined in Fig. \ref{fig:structures}.
			\label{tab:GD2Mag}
		\end{table*}

%
%
%
%

                \section{Results and Discussion}
	        \subsection{Structure and Bonding}

		\begin{figure}
			\includegraphics[width=375px]{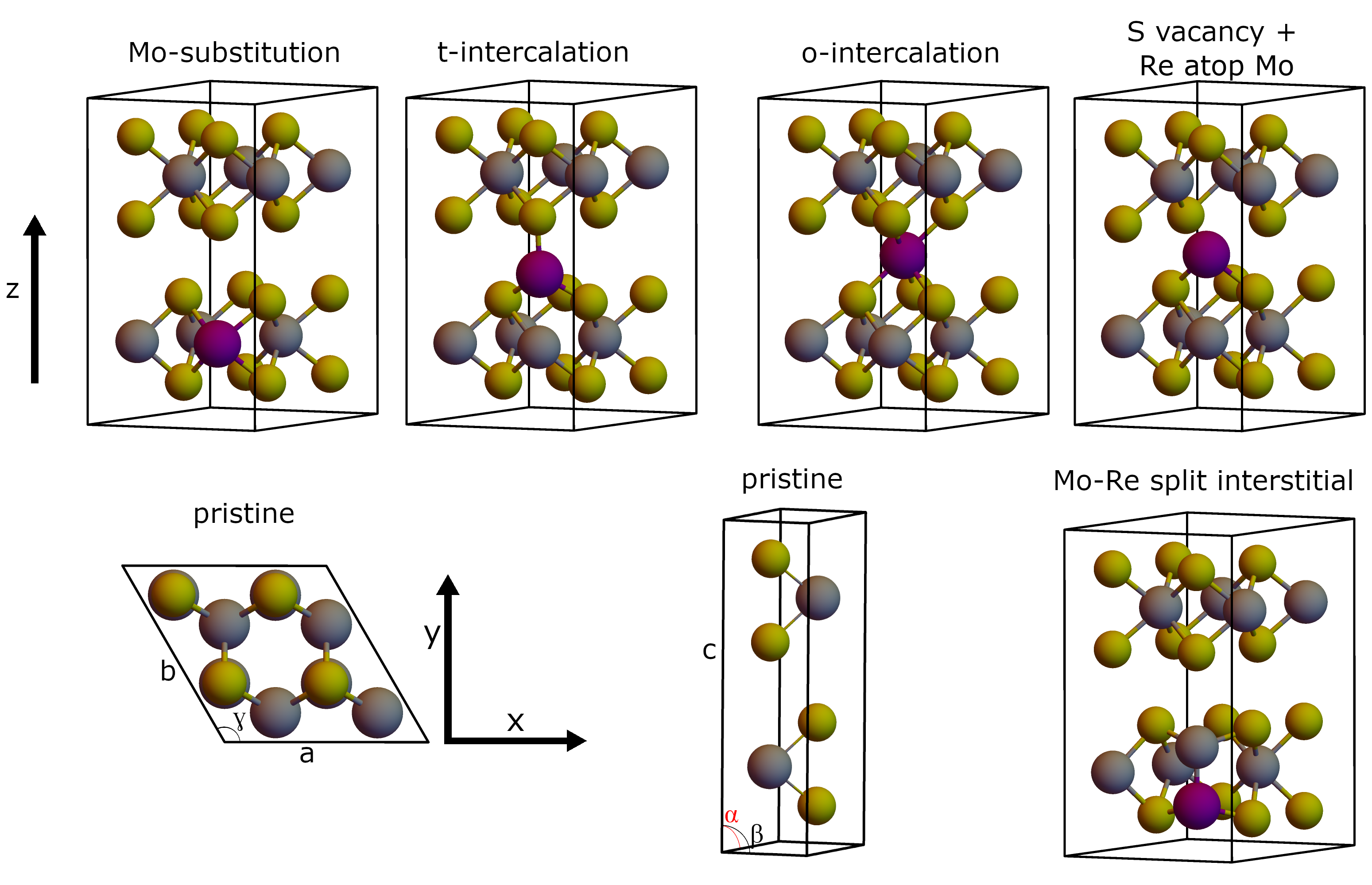}
			\caption{Re-doped MoS$_2$ structures found to be stable or metastable: Mo in grey, S in yellow, and Re in purple. O-intercalation and Mo-Re split-interstitial are metastable with respect to t-intercalation.}
			\label{fig:structures}
		\end{figure}

		\begin{figure}
			\includegraphics[width=450px]{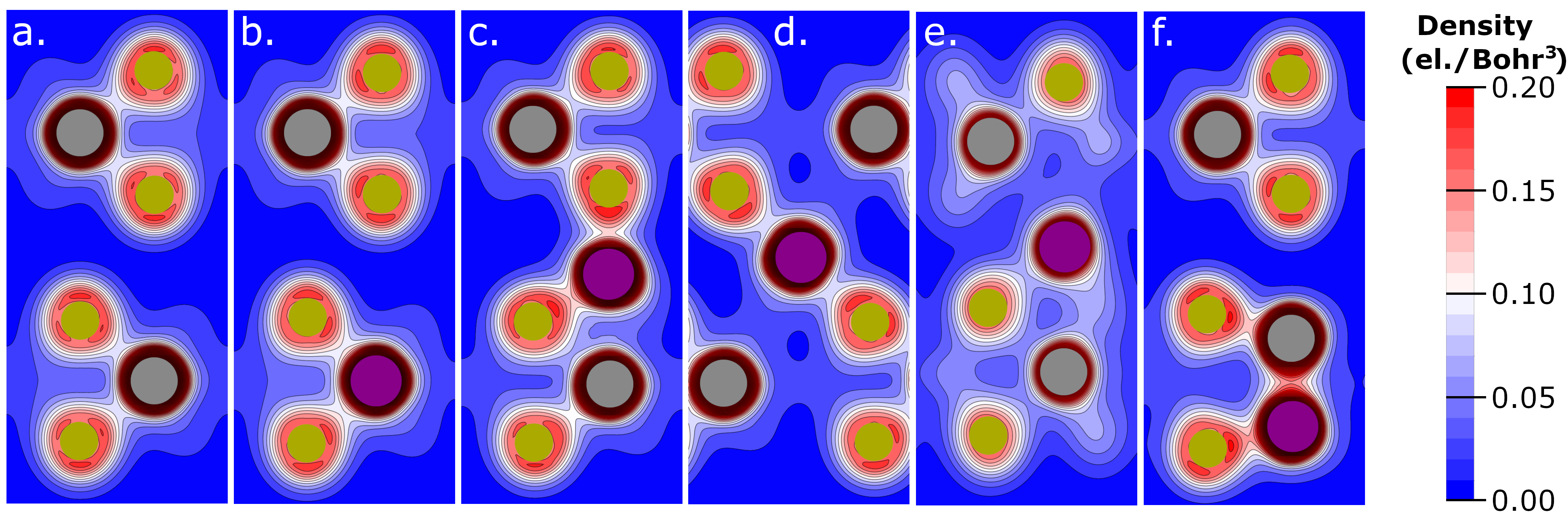}
			\caption{Electronic densities of MoS$_2$: a) pristine, b) Mo-substituted, c) t-intercalated, d) o-intercalated, e) S vacancy + Re atop Mo, and f) Mo-Re split-interstitial, in cross-sections of the $2\times2\times1$ supercell in an Mo-S plane close to the dopant. S atoms are shown in yellow, Mo atoms are in gray, and Re atoms are purple. Intercalated and S-vacancy structures show strong out-of-layer bonding. All structures besides the highly distorted S vacancy structure show densities in the Re-S bonding region comparable to Mo-S bonds.}
			\label{fig:Densities}
		\end{figure}

                We consider the following possible Re dopant sites: Mo substitution, S substitution, t-intercalation, octahedral (o-) intercalation, Mo-Re split interstitial (similar to the Mo-Mo split interstitial from \citet{Komsa}), bridge-site intercalation, and hollow-site intercalation. These structures were chosen as they are commonly considered in computational studies of MoS$_2$ doping.\cite{Ivanovskaya, Komsa} The bridge and hollow site structures (see Fig. S1) are unstable and relax to Mo-Re split interstitial and o-intercalation respectively, and are not considered further.
                The relaxed structures used are pictured in Fig. \ref{fig:structures}. We focused on three structures: undoped 2H-MoS$_2$, Mo-substituted, and tetrahedral-intercalation (t-intercalated). All structures retained a 2H structure and did not shift to, for example, the 1T phase, as has been obtained in some experimental studies.\cite{Xia, EnyashinRoute} There is one exception---the t-intercalated $3\times3\times1$ (with PBE+GD2) relaxed to a structure (Fig. S1) with the Mo atoms aligned above one another, in which the Re atom now has octahedral bonding to S atoms. This corresponds to aBa cBc stacking (using the convention in \citet{Song}), similar to the ``Min 2'' structure from sliding in \citet{Levita}. This structure is lower in energy by 0.02 eV per atom than if it had the typical stacking.

		The pristine structure has $a=b=3.190$ \AA,\ $c= 12.415$ \AA, $\alpha=\beta=90^\circ$ and $\gamma=120^\circ$, as shown in Fig. \ref{fig:structures}. For most of the computed structures, the lattice parameters match the pristine to within 0.5\%. There is a 1-2\%\ increase in $c$ in t-intercalation, consistent with a previous calculation \cite{Hallam} and similar to results for Ni \cite{Karkee} but considerably less than the increase in layer spacing for Li\cite{EnyashinLi}; the anomalous 3$\times$3$\times$1 structure shows a lesser increase in $c$. The split interstitial shows a 0.5-2.5\%\ increase in $a$ and $b$. The 2$\times$2$\times$1 and 2$\times$2$\times$2 S-substituted structures show a substantial change in the Re atom's location when relaxed---those structures are better described as S-vacancies with a Re dopant atop the Mo atom, as in Fig. \ref{fig:structures}. The charge densities show a pronounced effect from these atomic reorganizations (Fig. \ref{fig:Densities}). 
		Angles are within 0.05\%\ of the pristine with few exceptions: the t-intercalated 2$\times$2$\times$1, 2$\times$2$\times$2, and 3$\times$3$\times$1 supercells' $\alpha$ ($\beta$) are 0.4\%, 0.2\%, and 8.2\% larger (smaller) than the pristine $\alpha$ ($\beta$) respectively.
		In the Mo-substituted case, there is a noticeable local symmetry-breaking around the Re atom, leading to Re-S bonds in two groups measuring about 2.38 \AA\ and 2.41 \AA\ respectively at a 3$\times$3$\times$1 supercell. Due to the symmetry of the o-intercalation site, the six bond lengths occur in groups of two or three with bond length differences of around 0.005 \AA. Full structural information is found in Table S1, and complete structure files are available in the Supplementary Material. None of the doped structures has any exact symmetry, though some approximate symmetries remain.

		Analysis of the electronic density demonstrates that Re is able to form interlayer covalent bonds, as shown by significant electronic density between Re and S across layers in Fig. \ref{fig:Densities}(c,d): comparable to, or even larger than, the electronic density in Mo-S bonds. This happens in o- and t-intercalation, similar to our previous findings for Ni-doped MoS$_2$.\cite{Guerrero}. Mo-substitution is nearly indistinguishable from pristine in its electronic density, except that substituting increases the density in the S-bonding region in both layers. Re's presence in other cases reduces the electronic density in the region between two S atoms in the same layer. Intercalated Re also reduces electronic density between S atoms it is bonded to and their adjacent Mo atoms, suggesting a weakened bond. The non-intercalated doped structures however show only van der Waals interactions between layers.

	\subsection{Thermodynamic Stability}

		With the method we previously used for Ni-doped MoS$_2$ \cite{Guerrero}, we compute the phase diagram as a function of chemical potentials, marking the most stable structure under each condition.
		The relative stability of structures with different stoichiometries can be analyzed using the formation energy, $E_{\rm formation}$:
		\begin{equation}
			E_{\rm formation} = E_{\rm mixed} - \sum_{i}N_iE_{i,{\rm bulk}} - \sum_{i}N_i\mu_i
		\end{equation}
		$E_{\rm mixed}$ is the total DFT energy of the doped system. $\mu_i$ is the chemical potential of the species $i$, and $N_i$ is the number of that species in the mixed system. Stable bulk elemental energies $E_{i,{\rm bulk}}$ are computed from the lowest energy structures we computed from those found in the Materials Project\cite{MaterialsProject}, structures mp-129 (Mo), mp-557869 (S), and mp-8 (Re).

		Construction of the phase diagram requires the energies and composition of different phases of Re, Mo, and S compounds. We used the set of stable and metastable structures we found: Mo substitution, S substitution, t-intercalation, o-intercalation, and the Mo-Re split interstitial.
                O-intercalation, t-intercalation, and Mo-Re split interstitial have the same chemical composition and thus their energies are directly comparable. Of these three, t-intercalated consistently has a lower energy by 0.01--0.02 eV/atom. There is one exception: the $4\times4\times1$ split-interstitial is very slightly energetically favorable over either intercalated structure by 0.001 eV/atom. Full formation energies are given in Table S2.

\begin{figure}
			\includegraphics[width=300px]{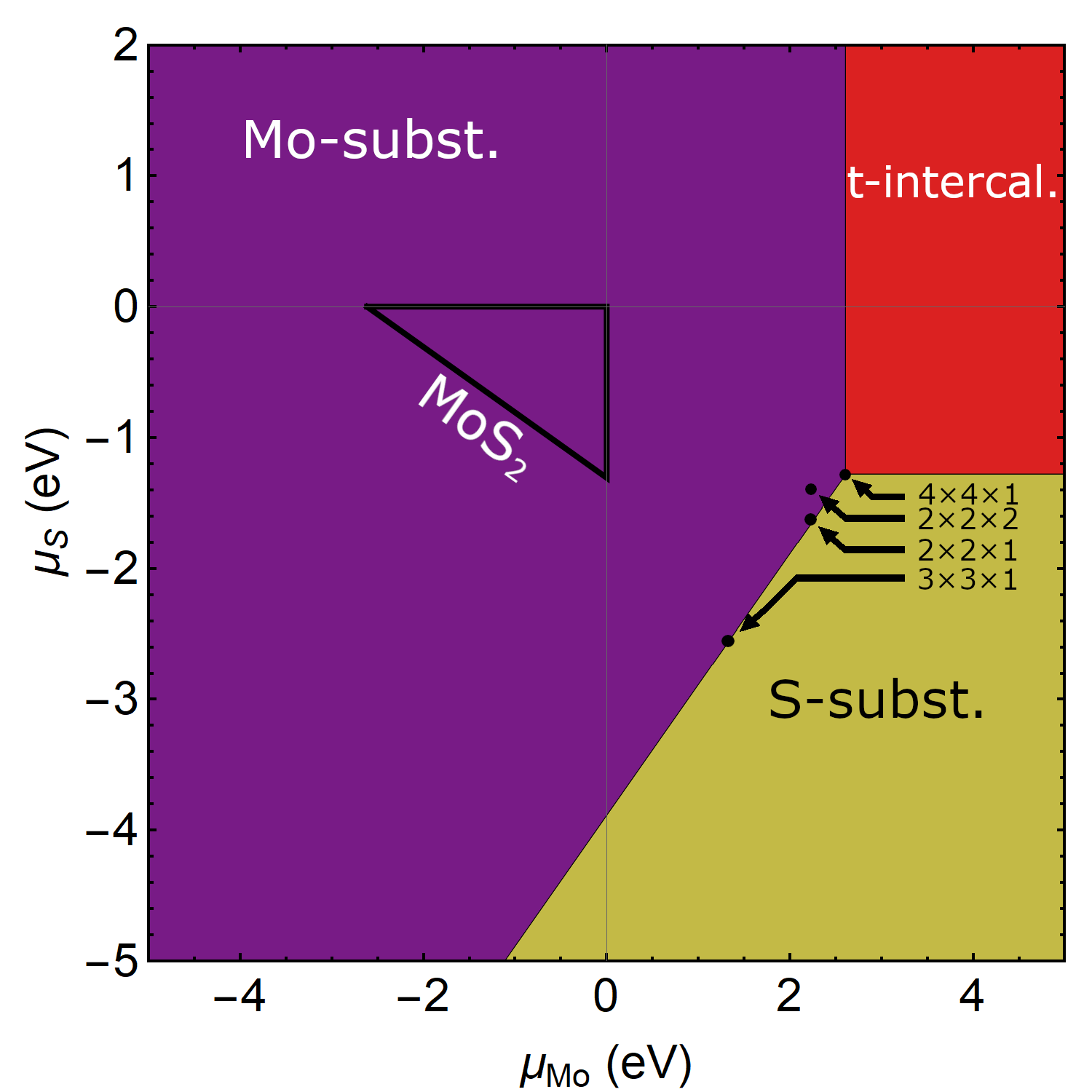}
			\caption{The $T=0$ K phase diagram for Re-doped MoS$_2$ in a $4\times4$ supercell. Only the Mo-substituted Re-doped MoS$_2$ falls within
                          the stability triangle (black) for MoS$_2$ (as in \citet{Guerrero}). The labeled dots show the location of the intersection point for other supercell sizes. The pristine MoS$_2$ line meets the axes at $\mu_{\rm S}=-1.308$ eV and $\mu_{\rm Mo}=-2.616$ eV. Structures considered include Mo-substituted, S-substituted, tetrahedral intercalation, octahedral intercalation, and the Mo-Re split interstitial. Note that for $4\times4\times1$ the Mo-Re split interstitial has essentially the same energy as t-intercalation.}
			\label{fig:phaseDiag}
		\end{figure}

		The $T$=0 phase diagram (Fig. \ref{fig:phaseDiag}) shows that Mo-substituted is the most stable structure, and the only one consistent with stability of pristine MoS$_2$. This result is consistent with experimental observation of Mo-substituted monolayers, \cite{LinDopant} which have a comparable geometry. While Re adatoms on monolayers are much less stable, bulk intercalation of Re has been inferred in experimental samples, \cite{Acikgoz} which can be due to out-of-equilibrium or higher-temperature synthesis processes. The picture is quite different from that for Ni-doping of MoS$_2$, in which only t-intercalation was compatible with pristine MoS$_2$ stability (except at the highest doping levels).\cite{Guerrero}
                
                Phase stability at higher temperature can be estimated in the harmonic approximation by calculating the lattice vibrations' contributions to the free energy \cite{Komsa} $F=E_0 + E_{\rm vib} - TS$. Added to the DFT energy $E_0$ are the vibrational energy $E_{\rm vib}$ (including zero-point energy)
                \begin{equation}
                  E_{\rm vib} = \sum_\lambda \left( n_B (\hbar \omega_\lambda, T) + \frac{1}{2} \right) \hbar \omega_{\lambda},
                \end{equation}
                and the temperature ($T$)-dependent entropy
                \begin{equation}
                  S(T)=k_B\sum_\lambda n_B(\hbar \omega_\lambda, T) \ln n_B(\hbar {\omega}_\lambda, T)
                \end{equation}
                with Bose-Einstein populations $n_B$ and phonon frequencies $\omega_\lambda$ for modes $\lambda$. We focus on $-TS$ which is predominantly responsible for the temperature dependence. \cite{Komsa} The lowest-frequency modes contribute most to the entropy term, namely the acoustic modes as well as shearing E$_{\rm 2g}^2$-like (35.2 cm$^{-1}$ in pristine) and layer-breathing B$_{\rm 2g}^2$-like (55.7 cm$^{-1}$ in pristine) modes. Compared to Mo substitution, t-intercalation has a 6--10 cm$^{-1}$ lower frequency in the E$_{\rm 2g}^2$-like mode but a 5--20 cm$^{-1}$ higher frequency (\textit{i.e.} stronger interactions) in the B$_{\rm 2g}^2$-like mode. Throughout most of the Brillouin Zone, the E$_{\rm 2g}^2$-like modes are the lowest in frequency, below even the acoustic modes, and will be significantly populated at room temperature ($k_B T \sim 200\ {\rm cm}^{-1}$). Therefore this mode can contribute considerably to the free energy. Since this mode is lower in energy for t-intercalation, it will contribute to a more negative entropy for t-intercalation vs. Mo substitution, and at high temperatures the free energy could favor t-intercalation, making it accessible even within the pristine stability triangle in Fig. \ref{fig:phaseDiag}. The influence of $E_{\rm vib}$ on the relative stability is small and not very temperature-dependent; all modes contribute and there is cancellation between the effects of some higher and some lower frequencies between the two structures.

	        \subsection{Raman spectra}

                To predict Raman spectra, we need both vibrational frequencies and intensities in general. Only the vibrational frequencies might be needed in some special cases if the intensities do not vary much between modes, if there are just a few Raman-active peaks which can be determined by symmetry, or if the intensities can be predicted from the pristine as a reference system by finding a simple relation between the doped and pristine modes.\cite{Sheremetyeva} As we will detail in this section, none of these situations applies in the case of Re-doped MoS$_2$, requiring a more general and complete approach.

                \subsubsection{Formalism}
        The standard approach to Raman spectra in DFPT \cite{Lazzeri} calculates atomic Raman tensors via electric-field and atomic-displacement perturbations, and then combines these with phonon eigenvectors to obtain the Raman intensities of phonon modes. The formalism is based on the Placzek approximation and a static approximation for the dielectric constant $\epsilon^\infty$, which is typically valid for sub-gap (non-resonant) incident light.
        \begin{equation}
        \label{eq:intensity}
        I_{i,s}^\nu \propto \left| \hat{e}_i \cdot A^{\nu} \cdot \hat{e}_s \right|^2 \frac{1}{\omega_\nu} \left( n_{\rm B} \left( \hbar \omega_\nu, T \right) + 1 \right)
        \end{equation}
        where $I_{i, s}$ is the intensity with polarization vectors $\hat{e}$ in incident direction $i$ and scattered direction $s$, $\nu$ is a mode index, $A_{\nu}$ is the mode Raman tensor, and $\omega_\nu$ is the phonon frequency. In this work, we consider unpolarized Raman, with an average over the polarizations for incident and scattered light.\cite{Porezag}
        The mode Raman tensor is calculated in terms of atomic Raman tensors $A^{k \gamma}_{lm}$ via
        \begin{eqnarray}
        \label{eq:raman_tensors}
        A^{\nu}_{l m} = \sum_{k \gamma} A^{k \gamma}_{l m} \frac{w^{\nu}_{k \gamma}}{\sqrt{M_\gamma}} \\
        A^{k \gamma}_{l m} = \frac{\partial^3 U^{el}}{\partial E_l \partial E_m \partial u_{k \gamma}} = \frac{\Omega}{4 \pi} \frac{\partial \epsilon^\infty_{lm}}{\partial u_{k \gamma}}
        \end{eqnarray}
        where $w^\nu$ is the displacement pattern of mode $\nu$; $k$, $l$, and $m$ are Cartesian directions, $\gamma$ is an atom index, $M_\gamma$ is the atomic mass, $\Omega$ is the unit cell volume, $U^{el}$ is the electronic energy, $u$ is an atomic displacement, and $E$ is the electric field. Clearly this approach is only meaningful when a finite $\epsilon^{\infty}$ can be defined. Metallic systems, i.e. those without a bandgap, cannot be calculated because of the divergence of the dielectric constant in this case. Practically speaking, any system treated with smearing in \texttt{Quantum ESPRESSO} falls in this category.

\begin{figure}
	  \includegraphics[width=450px]{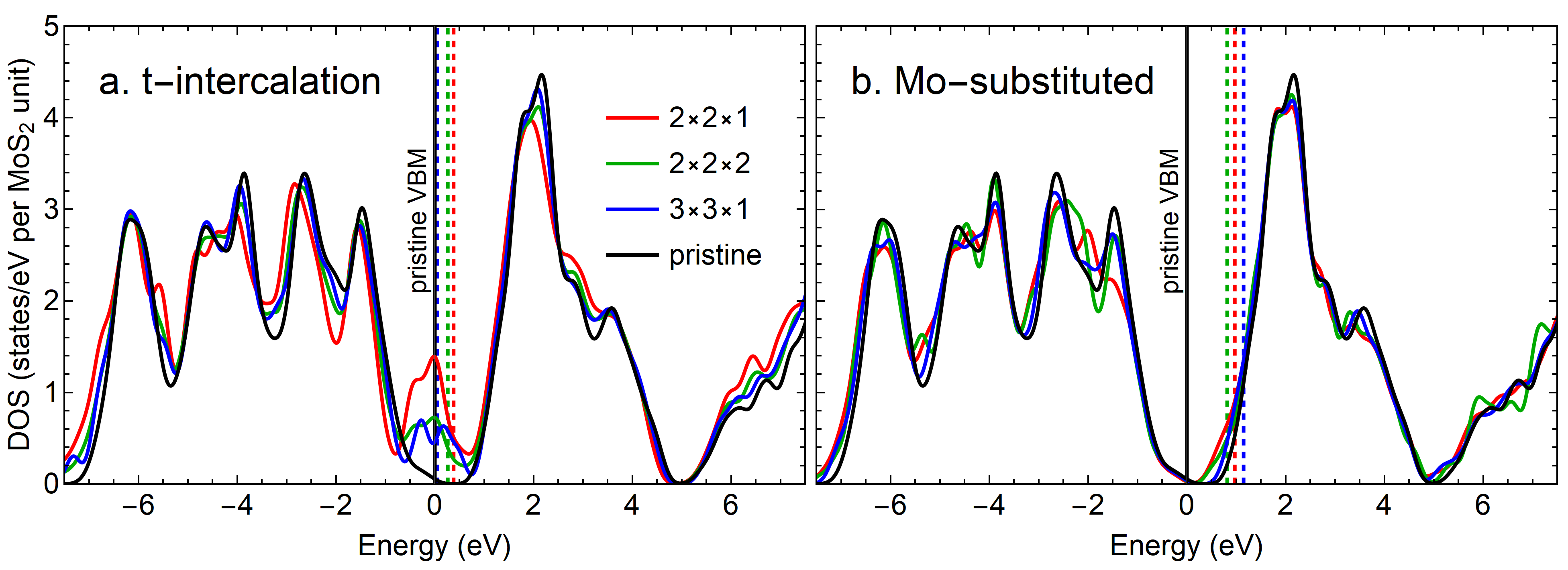}
	  \caption{Electronic density of states (DOS) for a) t-intercalated and b) Mo-substituted structures (using the LDA relaxed structures, as in the Raman calculations). Energies are aligned to the lowest-lying Mo states and referenced to the pristine valence band maximum. Vertical dashed lines indicate the Fermi energies. All doped structures have nonzero DOS at the Fermi level. Mo-substituted structures are clearly $n$-type with Fermi energies in the conduction band. A Gaussian broadening of 0.2 eV was used. }
	  \label{fig:EDOS}
	\end{figure}

        True metals generally do not have observable Raman intensity because light will be reflected rather than undergoing Raman scattering, due to a large $\epsilon^\infty$ at the incident frequency; exceptions are \citet{FeldmanDW} and a recent work on Cu surfaces.\cite{Denk} Doped semiconductors however do typically show Raman intensity, in particular for low doping concentration and therefore typically a low DOS at the Fermi level. The metallic nature in a small supercell calculation may be due to spurious formation of impurity bands. Moreover, $n$-type doping like Re-doped MoS$_2$ in a low-doping limit is expected to leave the bandgap intact and simply move the Fermi level into the conduction bands, as in \citet{Hallam} In this case, the dielectric constant remains finite at the optical frequencies of incident light for Raman experiments. Indeed this is exactly the situation sought for transparent conductors.\cite{TCO} In our calculation, all the t-intercalation and Mo-substituted structures have metallic DOS (Fig. \ref{fig:EDOS}). The Mo-substituted structures show a Fermi level in the conduction band which demonstrates $n$-type doping as expected from the literature, and the bandgap is still retained, particularly for lower-doped structures. For t-intercalation, we see significant modification of the DOS due to new in-gap states, rather than simply electron donation. 

        We also note that many experimental works have reported Raman scattering measurements from Re-doped MoS$_2$,\cite{Acikgoz, Hallam} so there should be a way to obtain these Raman tensors theoretically. Rather than use prohibitively large supercells and/or perform computationally intensive resonant Raman calculations (e.g. using time-dependent DFT or the Bethe-Salpeter equation \cite{YWang}), we develop an alternate approach based on the approximation that the atomic Raman tensors are similar to those of pristine MoS$_2$ and/or a similar reference system which has a well-defined gap and can be calculated in the usual way. We use directly calculated phonon displacement patterns and simply substitute the atomic Raman tensors from the reference system in Eq. \ref{eq:raman_tensors}. In practice, this is done by inserting the reference atomic Raman tensors from the undoped case into the files \texttt{Quantum ESPRESSO} uses to compute the Raman intensities. Individual Mo and S atoms are matched to their counterparts in a pristine structure. There are several plausible options for the dopant atom's contributions to the atomic Raman tensor: approximate as Mo, approximate as another dopant from a similar system (e.g. as Ni), or neglect as zero, which we test below.
        Our method could also be applied to materials with very large numbers of atoms per unit cell, such as amorphous materials; the phonon modes may be computed by classical force fields to save computational effort, while the tensor could use DFT methods. Atomic Raman tensors can be rotated when atoms appear with a different orientation of bonds, along the lines of Raman bond-polarizability models \cite{Wirtz}.

        Our scheme is related to a more limited approach taken in recent work \cite{Hashemi, Kou}, in which the Raman {\it intensities} from a reference system are used to calculate Raman spectra of a doped system, to avoid the need for explicit calculations on the doped system. It can be derived by projecting the doped mode ($\mu$) eigenvectors onto the pristine mode ($\nu$) eigenvectors (in a supercell corresponding to the doped system), giving coefficients $c_{\mu \nu}$:
        \begin{equation}
          \label{eq:overlap}
        c_{\mu \nu} = \sum_{k \gamma} u_{k \gamma}^{\mu,{\rm doped}} u_{k \gamma}^{\nu,{\rm pristine}}
        \end{equation}
The Raman tensors are then expanded in terms of those of the pristine reference system, $\tilde{A}$:
        \begin{eqnarray}
		\label{eq:Hashemi}
		|\hat{e}_i \cdot A^{\nu} \cdot \hat{e}_s|^2 = |\hat{e}_i \cdot \sum_{\mu} c_{\mu \nu} \tilde{A}^{\mu} \cdot \hat{e}_s|^2 \\
		= \sum_\mu c_{\mu \nu} \left( \hat{e}_i \cdot \tilde{A}^{\mu} \cdot \hat{e}_s \right) \sum_\sigma c_{\sigma \nu}^{*} \left( \hat{e}_{i} \cdot \tilde{A}^{\sigma *} \cdot \hat{e}_s \right) \\
		\label{eq:HashemiApprox}
		\approx \sum_\mu |c_{\mu \nu}|^2 | \hat{e}_i \cdot \tilde{A}^\mu \cdot \hat{e}_s|^2
	\end{eqnarray}
	Cross terms ($\mu \neq \sigma$) are neglected to reach Eq. \ref{eq:HashemiApprox}. The intensity $I$ can be written correspondingly in terms of the intensities $\tilde{I}$ of the reference system:
		\begin{eqnarray}
                  I_{i, s}^\nu = \sum_\mu |c_{\mu \nu}|^2 \tilde{I}_{i, s}^\mu
		\end{eqnarray}
        As in our method, only the phonon eigenvectors are needed for the system of interest, which could be computed by any means, including classical force fields.
        A key limitation is that working with only the scalar intensities means losing some information about the Raman response; in particular, interference effects (the cross terms in Eq. \ref{eq:HashemiApprox}) are neglected, which can be significant, in particular for the case of symmetry-breaking of a Raman-inactive mode into Raman-active modes.

        \subsubsection{Benchmarking with Ni-Doped MoS$_2$}

\begin{figure}
			\includegraphics[width=325px]{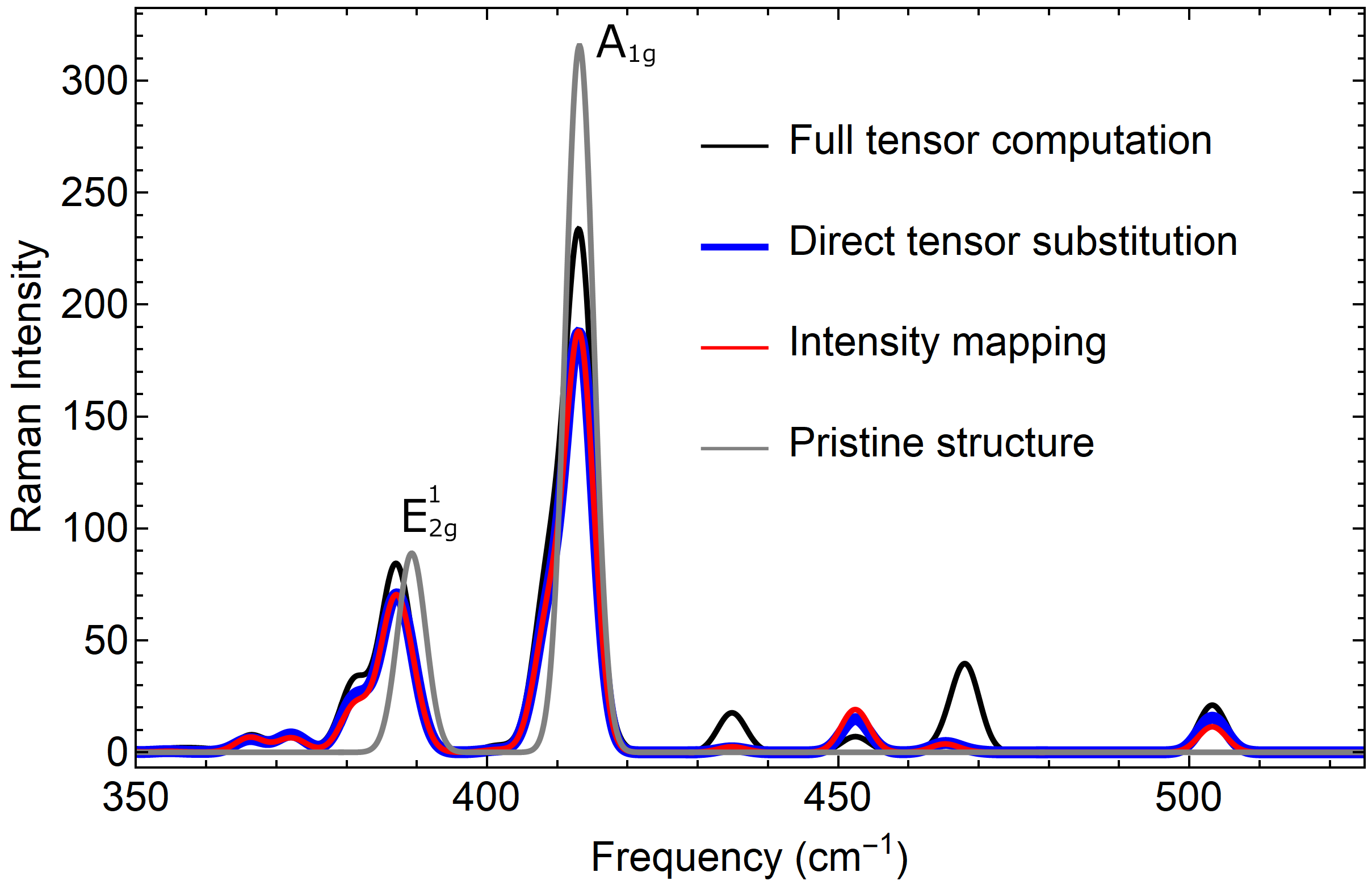}
			\caption{Raman spectra (\AA$^4$/amu per MoS$_2$ unit, unpolarized) of t-intercalated Ni-doped MoS$_2$, comparing intensity approximations to a full tensor computed using the Lazzeri method \cite{Lazzeri} which is not possible with Re-doped MoS$_2$. The irreducible representations for the pristine peaks are labelled. The tensor substitution is a slight improvement when compared to the intensity mapping \cite{Hashemi, Kou} method.}
			\label{fig:NiRaman}
		\end{figure}

	We tested our method, and the intensity mapping of Eq. \ref{eq:HashemiApprox}, on our previously computed $3\times3\times1$ supercell of t-intercalated Ni-doped MoS$_2$ \cite{Guerrero}, which does not require smearing and can be handled by standard methods. We use Raman tensors for Mo and S from pristine MoS$_2$, and initially set the Raman tensor of Ni to zero. The relative peak heights are 15\%\ below the full computation as shown in Fig. \ref{fig:NiRaman}, and the peak height ordering is preserved.
        Compared to the intensity-based scheme, we see that our Raman-tensor treatment makes small improvements near the pristine Raman-active peaks of Ni-doped MoS$_2$ (around 380 cm$^{-1}$ in Fig. \ref{fig:NiRaman}). Both methods significantly underestimate the intensity of some modes which are highly local to the Ni (435 cm$^{-1}$ and 470 cm$^{-1}$) due to lack of information about the distinctive contributions of these atoms, but are better at modes which contain both local components and pristine-like vibrations (452 cm$^{-1}$ and 502 cm$^{-1}$).

\begin{figure}
			\includegraphics[width=425px]{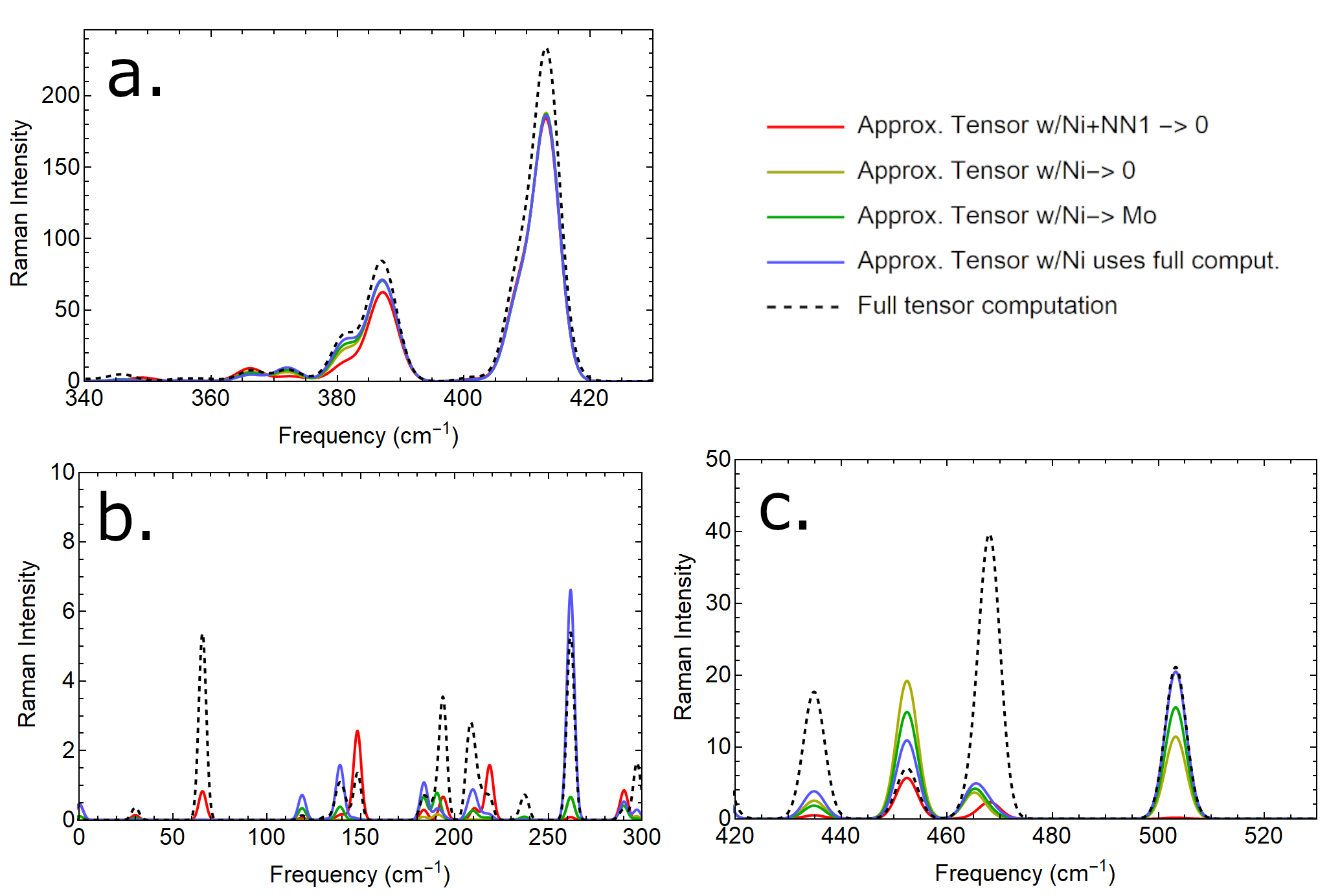}
			\caption{Comparison of the Ni-doped spectra (unpolarized) for different approximations of the Ni atomic Raman tensor across three ranges in \AA$^4$/amu per MoS$_2$ unit (note the different scale for the three plots). The Mo and S atoms use the same tensor as the pristine. For the red line, the Ni and its nearest neighbor (NN) S atoms' contributions to the Raman tensor have been set to 0. For the rest, the Ni atom's contribution only has been set to 0 (yellow), the same as Mo in pristine (green), and the same as the full computation (blue). The pristine-active modes (a) show the least amount of change, except on the shoulder of the E$_{\rm 2g}^1$ peak. For the high-frequency modes (c), the 435 cm$^{-1}$ and 470 cm$^{-1}$ peaks are highly localized to the Ni and are not described well. At 505 cm$^{-1}$, the peak is entirely localized near the Ni and its NN, as evidenced by the lack of intensity when those contributions are removed.}
			\label{fig:NiRamanTensor}
	\end{figure}

        We further found that by comparing structures with different choices for replaced Raman tensors against the pristine result, we can identify which peaks in the spectra are due to geometry changes rather than new Ni-bond related activity. In this case Ni introduces shifts to the active peaks, new peaks related to new modes local to the Ni, and activations of existing MoS$_2$ modes that were previously inactive by symmetry \cite{Guerrero}. Our Raman tensor approximation method is versatile when distinguishing changes in the intensities as being caused by lattice distortions or charge transfer from doping.
        By setting to zero specific atomic components (e.g. those of the dopant and its nearest neighbor) we are able to analyze the effect of different atoms in the Raman scattering intensity.
	In Fig. \ref{fig:NiRamanTensor}, we test different approximation schemes for the Ni component of the atomic Raman tensor. Setting Ni and its nearest neighbors' (NN) contributions to zero is the most reductive approximation and is thus furthest from accurate---this is clearest in the muted shoulder of the E$_{\rm 1g}$ peak. It is still useful, however, in interpreting the nature of modes such as the 505 cm$^{-1}$ peak. The lack of intensity in the NN line means that activity of this peak is highly localized to the Ni atom. The approximation works well near the pristine-active peaks, but fails whenever the activity is localized around the Ni atom. The fact that the peak around 150 cm$^{-1}$ has too \textit{large} an intensity when these contributions are set to zero underscores the importance of interference effects in the Raman intensities, as captured by our approach. We can see the small effect of Ni-doping on the Raman tensors of Mo and S (neglected in our scheme) by examining the small difference between the results of the full computation and the results of the approximation in which Mo and S tensors are from pristine and the Ni tensor is from the full computation. We see better agreement with the full computation when the Ni tensor is approximated as Mo than when it is set to zero; this ability to control contributions from the dopant atom is an advantage of our scheme over one using just intensities. Since altering just the Ni dopant's contribution does not change the pristine-active region much, we expect that this method can predict the Re-doped structures' spectra with little loss in accuracy in this frequency regime.

		\begin{figure}
			\includegraphics[width=425px]{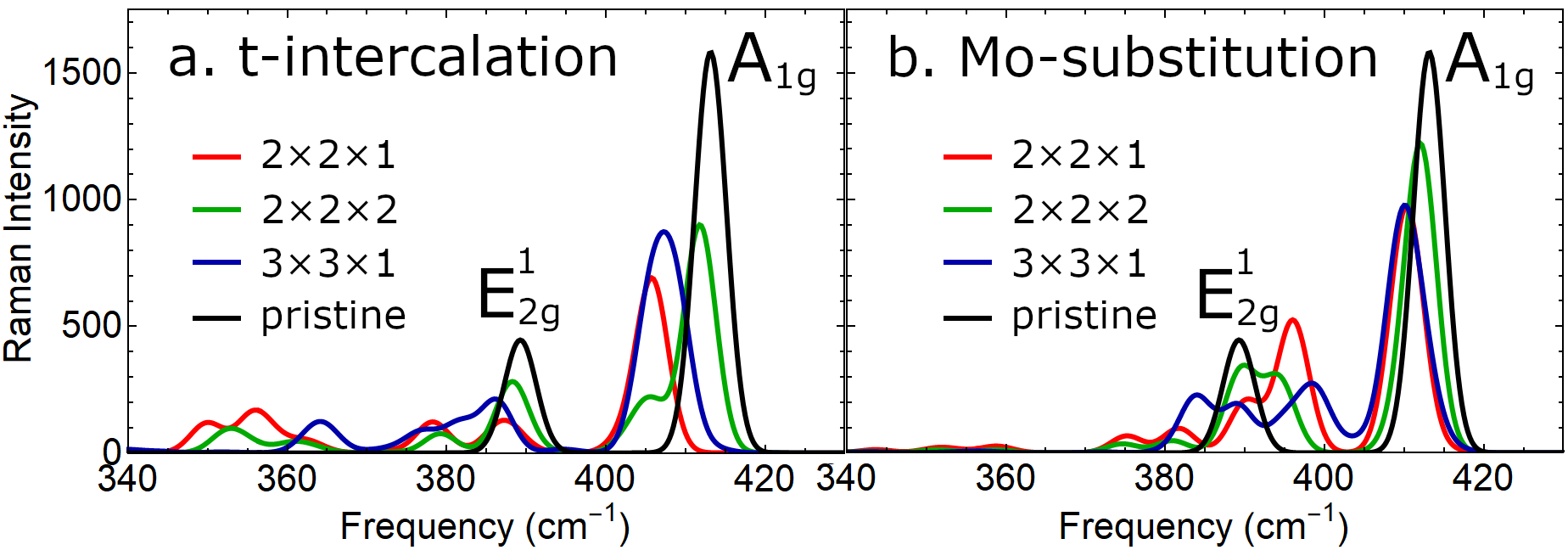}
			\caption{Raman spectra (\AA$^4$/amu per MoS$_2$ unit, unpolarized) of t-intercalated and Mo-substituted Re-doped MoS$_2$ as computed by DFPT with substitution of the Raman tensor by those in a pristine computation (Mo and Ni replace the Re atom's contribution in the Mo-substituted and t-intercalated respectively). The irreducible representations for the pristine peaks are labelled. The direction of the peak shifts is generally preserved among the supercell sizes used: redshifts in both peaks when t-intercalated and blue- and redshifts in the E$_{\rm 2g}^1$ and A$_{\rm 1g}$ peaks respectively while Mo-substituted. Gaussian broadening of 2 cm$^{-1}$ was used.}
			\label{fig:Raman}
		\end{figure}

		\begin{table*}[t]
			\setlength{\tabcolsep}{6pt}
			\renewcommand{\arraystretch}{1.5}
			\begin{tabular}{| c | c | c | c |}
				 \hline
				 Dopant Site & Supercell & A$_{\rm 1g}$ (cm$^{-1}$) & E$_{\rm 2g}^1$ (cm$^{-1}$) \\
				\hline \hline
				 t-intercal. & $2\times2\times1$ & -7.4 & -2.0 \\
				 \hline
				 t-intercal. & $2\times2\times2$ & -1.3 & -1.0 \\
				 \hline
				 t-intercal. & $3\times3\times1$ & -5.9 & -3.2 \\
				 \hline
				 Mo-subst. & $2\times2\times1$ & -2.8 & +6.8 \\
				 \hline
				 Mo-subst. & $2\times2\times2$ & -1.1 & +0.7, +4.5 \\
				 \hline
				 Mo-subst. & $3\times3\times1$ & -3.1 & -5.3, -0.34, +9.1\textsuperscript{\emph{a}} \\
				 \hline
			\end{tabular}
			\caption{Vibrational frequency shifts for t-intercalated and Mo-substituted structures with respect to undoped MoS$_2$.}
			\label{tab:RamFreqShift}
                        \textsuperscript{\emph{a}} Appears near E$_{\rm 2g}^1$ but actually more related to A$_{\rm 1g}$ in character.
		\end{table*}

                \subsubsection{Application to Re-Doped MoS$_2$}
                
\begin{figure}
  \includegraphics[width=400px]{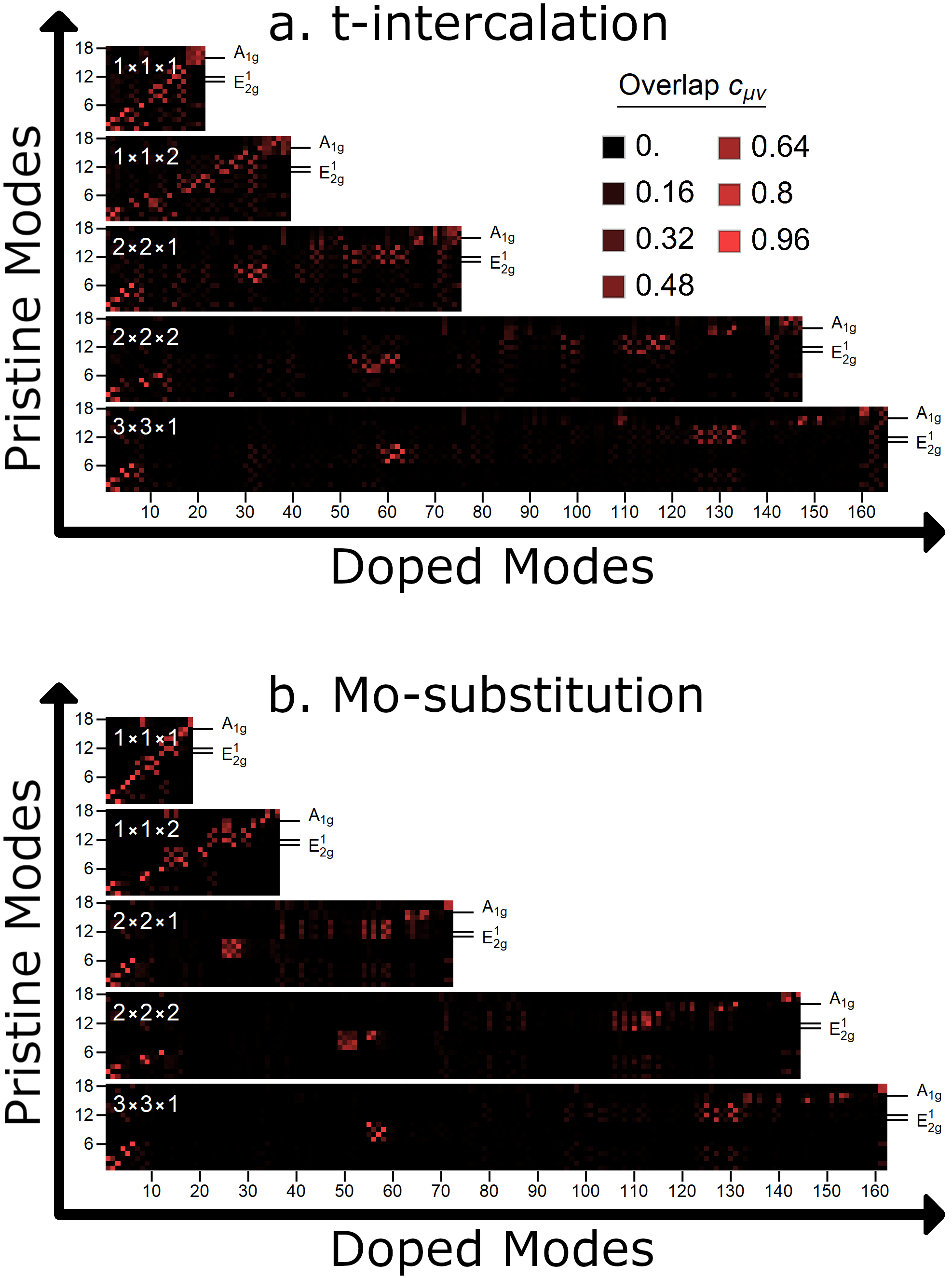}
  \caption{Elements of the matrices $c_{\mu \nu}$ representing the overlap between displacement patterns $u$ in Re-doped and pristine structures, per Eq. \ref{eq:overlap}. Modes are ordered by increasing frequency and the Raman-active modes of the pristine MoS$_2$ are marked.}
  \label{fig:Overlap}
\end{figure}

		Now we turn to our main interest, the Re-doped case. As for Ni-doped MoS$_2$, the doped structure closely resembles the pristine structure, making our approximation reasonable. In Mo substitution, we approximate the Re tensor with Mo, and in t-intercalation, we approximate Re by the Ni tensor from t-intercalation in a $3\times3\times1$ supercell, as this is a transition metal in the correct bonding geometry.
                Raman spectra in the pristine-active frequency regime are plotted in Fig. \ref{fig:Raman}. Complete Raman spectra and vibrational density of states (VDOS) are provided in Figs. S2-S4. Phonon eigenvectors and dynamical matrix files are also in the Supplementary Material. We find a consistent pattern among the Mo-substituted and t-intercalated Re-doped Raman spectra for different supercells we computed, as summarized in Table \ref{tab:RamFreqShift}: there is a redshift of both peaks in the t-intercalated case and blue- (E$_{\rm 2g}^1$) and red- (A$_{\rm 1g}$) shifts in the Mo-substituted case. The t-intercalated shifts are consistent with some experimental reports on multilayers \cite{AlDulaimi, Ghoshal, Xia} and monolayers \cite{LiShiSheng}, though the Raman spectra were not typically studied in much detail and A$_{\rm 1g}$ shifts were not resolved except in \citet{AlDulaimi} This work argued that a red shift of E$_{\rm 2g}$ was a sign of Mo substitution in their sample, and the effect of a heavier mass of Re than Mo,\cite{AlDulaimi} but our results show that this interpretation is not correct since Mo substitution in fact blueshifts this peak, with bonding effects outweighting the mass effects. The agreement between our intercalated results and monolayer experiments suggests adatoms on those monolayer samples. The changes in the spectra are not purely described by single-mode frequency shifts. Fig. \ref{fig:Overlap} shows the overlap between doped and undoped modes---the presence of large ``off-diagonal'' components shows that there is not a single doped mode that can be used to describe the shifts in frequencies of the pristine mode. Activations of new modes are also possible (since the doped structures no longer have any exact symmetry), as seen by comparing the VDOS of the pristine and doped plots in Fig. S4.
Another experiment on monolayer samples believed to be Mo-substituted \cite{GaoSubst} showed shifts consistent with our calculations for Mo substitution.
The bulk Raman spectra of Hallam \textit{et al.} \cite{Hallam} by contrast do not have the resolution to show any doping-induced shifts. Different synthesis methods and conditions can cause different doping sites (or distributions of sites) in their samples. It is worth noting that Z-contrast imaging \cite{LinDopant} is often used to locate the Re dopant, but the basal (in-plane) position of the t-intercalated dopant is the same as the Mo or S site, depending on which side it is viewed from, and so this method cannot give a conclusive answer. We can compare our calculated shifts to our previous Raman calculations for Ni-doped MoS$_2$: in that case, both Mo-substitution and t-intercalation caused red shifts for both of the main Raman peaks, making it harder to use for experimental identification. In both cases, t-intercalation induces extra peaks around 450-500 cm$^{-1}$, but whereas Re Mo-substitution induces few new peaks below 400 cm$^{-1}$, there are many significant new peaks in that range for Ni.

Inspection of the phonon modes that contribute to the Re-doped spectra show that the $3 \times 3 \times 1$ Mo-substituted structure's peak near 395 cm$^{-1}$ is (surprisingly) more related to A$_{\rm 1g}$ rather than E$_{\rm 2g}$. This means that the A$_{\rm 1g}$ mode splits and both peaks are redshifted. The nature of the 395 cm$^{-1}$ peak would have experimental signatures in polarized Raman. Redshifts in the t-intercalated case are supplemented by activations and mixing with the forbidden B$_{\rm 2u}$ which has a slightly lower frequency. High-frequency vibrations of the Re atom exist above the A$_{\rm 1g}$ peak (Fig. S4), but the (out-of-plane) A$_{\rm 1g}$-like vibrations are lower in frequency than the pristine---despite the increase in out-of-plane stiffness found in our calculations \cite{Acikgoz}. Comparing Mo-S interatomic force constants, those nearby the Re are lower in magnitude by about 25\% than those farther from Re in the $z$-direction. This is consistent with the observation that Re-doping lowers the A$_{\rm 1g}$ frequency in both cases. Magnitudes of force constants in the $x$- and $z$-directions are lower for t-intercalation than pristine by roughly 10\%, but are not appreciably changed for Mo-substituted, which is consistent with the different effects on the (in-plane) E$_{\rm 2g}$ frequency between the two dopant sites. The reduction in force constant is also consistent with the reduced Mo-S electronic density near Re for t-intercalation, indicating a weakening of the covalent bond (Fig. \ref{fig:Densities}). We note the complexity of the interactions here between the dopant and different modes -- quite different from the typical idea that doping simply increases or decreases bond strength and thus gives small shifts in modes that essentially retain their pristine character \cite{Iqbal}, or that shifts are due simply to the effect of the different mass of dopant atoms.

For analysis of Raman spectra, under some circumstances, just the frequencies can be sufficient for comparison to experiment, such as the case of pristine MoS$_2$, where there are only a few vibrational modes and the Raman-active ones can be identified by symmetry. Another case is Ag-intercalated few-layer MoS$_2$, in which it was found that the overlaps in displacement patterns can be used to find a one-to-one correspondence between the undoped and doped cases for interlayer breathing- and shearing-type modes, and therefore just the frequencies can be used to analyze how peaks shift.\cite{Sheremetyeva} However, in our case we find a rich set of vibrational modes for the doped structures, many of which have low intensities and little contribution to the Raman spectra. This effect can be seen clearly by comparison of the Raman spectra and the VDOS (Fig. S4). Therefore calculation of the intensities is essential to predict the peaks in the spectrum. The presence of the dopant can lead to increases in intensity (e.g. activation of an inactive mode by symmetry-breaking) or decreases in intensity (due to destructive interference between two Raman-active modes, as seen for the Ni-doped mode at 150 cm$^{-1}$ in Fig. 6). Moreover, there is rarely a single doped mode that corresponds to the pristine mode, or vice versa, as seen in Fig. \ref{fig:Overlap}. To determine the peaks in the Raman spectrum requires including the contributions of the Raman-active pristine modes to various doped modes.

	\section{Conclusion}
		We have computed properties of bulk Re-doped MoS$_2$ in different dopant configurations. We found that when intercalated, Re forms covalent interlayer bonds, and the tetrahedral geometry is most stable. In some cases, S-substituted structures rearrange into an S vacancy and Re adatom on the opposite layer. Mo-substitution is the most thermodynamically stable at equilibrium, though t-intercalation may be favorable at high temperature, and in general other structures can be formed out of equilibrium. These considerations of structure, bonding, and thermodynamics provide the basis for investigation of other properties such as friction and elasticity \cite{Acikgoz}.

		We developed a new efficient method to calculate the Raman spectra of metallic doped systems with an accurate approximation benchmarked on our previously calculated Ni-doped Raman spectra. The approach is easily performed in \texttt{Quantum ESPRESSO} by substituting atomic Raman tensors from a reference system. We find the t-intercalated Raman spectra shifts vs. the pristine in ways that are consistent with experimental literature on Re-doped MoS$_2$. The effects of doping on the E$_{\rm 2g}^1$ peak are more pronounced than on the A$_{\rm 1g}$ peak, with respect to both the frequency shifts and the degree to which they split into multiple related peaks.
The direction of the E$_{\rm 2g}^1$ peak shift can be used to identify the dopant site in experimental samples---Mo-substitution shows blueshifts (and sometimes a smaller redshift as well) while t-intercalated shows redshifts. These features we have identified provide an avenue to distinguishing the doping site in bulk phases which have been a challenge to identify experimentally \cite{Tedstone, Vazirisereshk} and may be different from the well-studied monolayer.
                Importantly, we have shown that the relationship between doping and peak shifts is complex, as also elucidated in our work on Ni-doped MoS$_2$ \cite{Guerrero}. The effect is not simply an overall strengthening or weakening of bond strengths by a dopant, but rather there are local effects around Ni, different effects on different bonds, interaction between different modes, and emergence of new dopant-related modes. In particular, a mode with frequency similar to E$^1_{\rm 2g}$ turned out to be more related to A$_{\rm 1g}$. There is no simple one-to-one correspondence between doped and pristine modes. Continued \textit{ab initio} study is needed to understand all the complexities of how dopants affect Raman spectra of 2D materials, and enable accurate interpretation of experimental spectra in terms of the electronic and structural effects of dopants.

\begin{suppinfo}
Structural information, formation energies, density of states, and full frequency range of Raman and vibrational density of states (PDF); XSF file-format structure files (ZIP), AXSF file-format phonon eigenvectors (ZIP), and dynamical matrix files from Quantum ESPRESSO (ZIP).
\end{suppinfo}

\begin{acknowledgement}
We acknowledge helpful discussions with Mehmet Z. Baykara. This work was supported by the Merced Nanomaterials Center for Energy and Sensing (MACES) via the National Aeronautics and Space Administration (NASA) Grants No. NNX15AQ01 and NNH18ZHA008CMIROG6R, and also by a Cottrell Scholar Award from the Research Corporation for Science Advancement, No. 26921. Computational resources were provided by the Multi-Environment Computer for Exploration and Discovery (MERCED) cluster at UC Merced, funded by National Science Foundation Grant No. ACI-1429783, and by the National Energy Research Scientific Computing Center (NERSC), a U.S. Department of Energy Office of Science User Facility operated under Contract No. DE-AC02-05CH11231.
\end{acknowledgement}

\bibliography{ReDopedRaman}

\section*{TOC Graphic}
\label{For Table of Contents Only}
\centering
\includegraphics[]{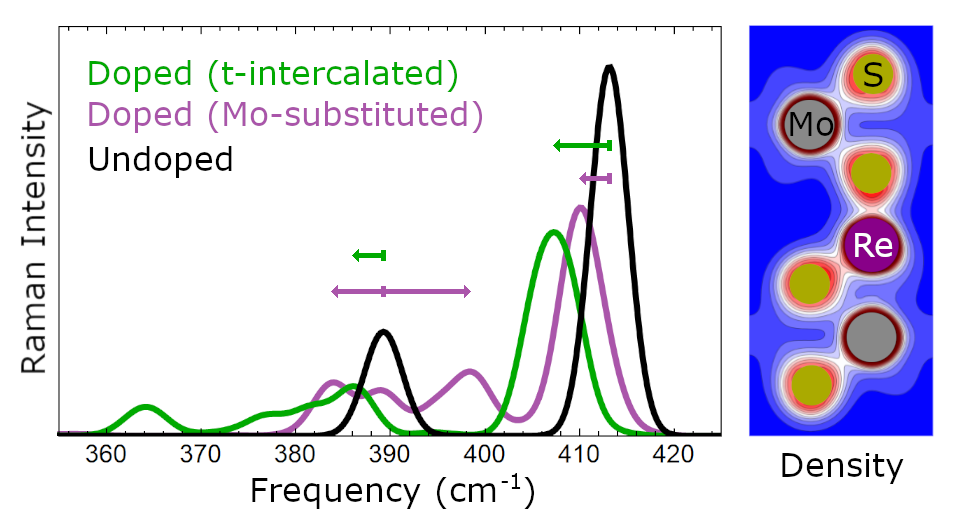}

\providecommand{\latin}[1]{#1}
\makeatletter
\providecommand{\doi}
  {\begingroup\let\do\@makeother\dospecials
  \catcode`\{=1 \catcode`\}=2 \doi@aux}
\providecommand{\doi@aux}[1]{\endgroup\texttt{#1}}
\makeatother
\providecommand*\mcitethebibliography{\thebibliography}
\csname @ifundefined\endcsname{endmcitethebibliography}
  {\let\endmcitethebibliography\endthebibliography}{}

\end{document}